\documentclass[12pt]{article}
\usepackage{amsmath,amssymb,amsthm,amssymb}

\title{On the problem of completeness of QM: von Neumann against
Einstein, Podolsky, and Rosen}
\author{Andrei Yu. Khrennikov\\
International Center for Mathematical
Modelling \\ in Physics and Cognitive Sciences,\\
University of V\"axj\"o, S-35195, Sweden\\
Email:Andrei.Khrennikov@msi.vxu.se}
\begin{document}

\maketitle

\abstract{We performed a comparative analysis of the arguments of
Einstein, Podolsky and Rosen -- EPR, 1935: \cite{EPR} (against the
completeness of QM) and the theoretical formalism of QM (due to
von Neumann, 1932: \cite{VN}). We found that the EPR
considerations do not match at all with the von Neumann's theory.
Thus EPR did not criticize the real theoretical model of QM. The
root of EPR's paradoxical conclusion on incompleteness of QM is
the misuse  of von Neumann's projection postulate. EPR applied
this postulate to observables with degenerate spectra (which is
totally forbidden by the axiomatics of QM).}

\section{Introduction}

During last 70 years completeness of QM and "quantum nonlocality"
have been the most intriguing problems in quantum foundations.
Since recently ideas on impossibility to provide a deterministic
description of reality (to introduce "hidden variables") and on
"quantum nonlocality" diffused outside of physics, e.g., to
philosophy, cognitive science, genetics, psychology and even
parapsychology, these problems became of the really
multi-disciplinary character.

To understand correctly such fundamental problems,  it is
extremely important to read carefully original sources. And I
would like to point out that the situation for mentioned problems
is astonishing. Although the original paper of Einstein, Podolsky
and Rosen -- EPR, 1935: \cite{EPR} is widely cited, it seems that
not so many people read it carefully (if at all!).

\subsection{Misuse of the von Neumann's projection postulate in
 EPR's argument}

In the present article I perform a careful analysis of the EPR
argument on the problem of completeness of QM. The conclusion of
such analysis is that EPR simply made a mistake in consideration
of the process of reduction of the wave function. The root of
EPR's paradoxical conclusion on incompleteness of QM is the {\it
misuse of von Neumann's projection postulate.} EPR applied this
postulate to observables with degenerate spectra (which is totally
forbidden by the axiomatics of QM, von Neumann, 1932: \cite{VN}).

I think that understanding of the real root of the EPR-paradox is
extremely important for quantum foundations. I hope that the
present paper would essentially clarify this problem.

\subsection{Copenhagen and V\"axj\"o interpretations of QM}

After publication of this preprint I was accused by some my
colleagues that I "changed the camp" and I took the side of the
orthodox Copenhagen community, e.g. "By reading your previous
papers one had an impression that you believed that QT should be
completed by some microscopic field theory. It seems quite strange
that you are using now the axiomatic approach of von Neumann, who
incorrectly claimed to prove the completeness of QT, in order to
prove the incorrectness of  EPR arguments." Therefore I should
explain from the very beginning the aim of this publication and my
own position.

My own position is the same as before, see e.g. \cite{VI}. I do
not think that the Copenhagen interpretation is the correct
interpretation of QM. I recall the main distinguishing features of
the Copenhagen interpretation:

\medskip

{\bf CH1}: Any state of an {\it individual}  physical system is
described by a wave function $\psi;$

\medskip

{\bf CH1}: The state of a system after measurement is determined
by the projection postulate.

\medskip

I think that the correct interpretation is so called statistical
interpretation. Recently it also becomes known as the {\it
V\"axj\"o interpretation}, see papers in \cite{R17}--
\cite{R19}.\footnote{The terminology "statistical interpretation"
which was elaborated and advocated by L. Ballentine \cite{BL},
\cite{BL1} is sometimes misleading, because some people using the
Copenhagen interpretation are also sure that they use "statistical
interpretation", since they use Born's rule. It became evident for
me in a series of discussions with Slava Belavkin who definitely
uses the Copenhagen interpretation, but at the same time he is
sure that it is "statistical interpretation."}

I recall the main distinguishing features of the V\"axj\"o
interpretation:

\medskip

{\bf VXU1}: A wave function $\psi$ is not an attribute of a single
physical system (e.g. electron).  A wave function $\psi$ (as well
as a density matrix $\rho)$ describe an ensemble of identically
prepared physical system. \footnote{Thus, opposite to the
Copenhagen interpretation, by the V\"axj\"o interpretation there
is no difference between "pure" and "mixed" quantum states. Both
types of states describes "subquantum mixtures".}

\medskip

{\bf VXU2}: The projection postulate determines not the state of a
system (after the corresponding measurement), but the probability
distribution of an ensemble of (output-)systems.

\medskip

This interpretation was supported by Einstein. In fact, article
\cite{EPR} was written to support this interpretation via proving
inconsistency of the Copenhagen interpretation.

I am definitely on Einstein's side regarding the interpretation of
QM. However, I think that arguments used to criticize opponents
should be perfectly rigorous. Otherwise such arguments might
induce even more misunderstanding.  The aim of my paper is to show
that, in spite of good wish of EPR, their arguments were not
rigorous. They misused the projection postulate. As a consequence,
the EPR paper became the source of

\medskip

a) {\it naive realism} -- an attempt to ignore the role of
measurement devices and assign values of e.g. two incompatible
observables to the same system;

\medskip

b) {\it quantum nonlocality.}

\medskip

At the first sight, the b) is surprising. EPR considered it as an
absurd alternative to a). Nevertheless, quantum majority took this
idea seriously. And we shall see that it was motivated by the very
structure of the EPR-arguments.

Thus my reply to supporters of the V\"axj\"o interpretation is
that even  the orthodox Copenhagen interpretation is better than
naive realism.

In this paper I shall show that one might work in the orthodox
Copenhagen framework {\it without quantum nonlocality!} To proceed
in this way, one should apply the projection postulate as it was
proposed by von Neumann.

Thus the main aim of this paper is to liberate the the orthodox
Copenhagen interpretation from the monster of quantum nonlocality.
It would be much easier to find common points between supporters
of the {\it local Copenhagen interpretation} and the V\"axj\"o
interpretation.

Concerning the critique of my colleagues from the V\"axj\"o side.
I agree that if one starts from the very beginning with the
statistical interpretation (the V\"axj\"o interpretation), one can
easily resolve the EPR paradox, see e.g. the excellent paper of
Kupczynski \cite{MK}. But it was not the aim of EPR! They used
their arguments for another purpose -- to destroy the  Copenhagen
interpretation.

\subsection{Von Neumann's postulate and L\"uders postulate}

The main point of this paper is that EPR applied the projection
postulate to operators with degenerate spectrum. Even if one takes
for a single system an operator with nondegenerate spectrum $A$,
e.g., spin, then by considering a pair of particles one should
realize this operator in the tensor product as $A\otimes I.$ So,
the latter has degenerate spectrum. Von Neumann's \cite{VN}
projection postulate is unapplicable in such a case. The postulate
which was used by EPR became later formalized by L\"uders, see
\cite{KHRV} for discussion.

My colleagues became angry again. This time I was attacked from
both sides, both from the Copenhagen and anti-Copenhagen.
Surprisingly both groups have the same viewpoint to the projection
postulate.

Copenhagen: "Whether or not it follows from von Neumanns'
axiomatization is irrelevant. There argument does follow from the
axiomatization adopted by all working physicists, still today. And
I suppose the argument had been used before EPR, they did not
invent it. When you have a composite system and you measure one
part of it, the joint state is projected into the subspace
obtained by taking the tensor product of the eigenspace of the
observable you have measured on one of the components, with the
whole of the second space. Are you saying that all books on
quantum information should be thrown away because this axiom was
not written down by von Neumann? Read any book on quantum
information eg Nielsen and Chuang."

Anti-Copenhagen:  "The thousands of physicists reading the EPR
paper did not object the reduction argument because they used it
in the same way.  Note that presently nearly all people working in
the field of quantum information are using the projection
postulate similarly as it was used by EPR."

First, I reply to the supporter of Copenhagen. Well, physicists
ignores von Neumann's distinction between operators with
degenerate and nondegenerate spectra in application of the
projection postulate. But they pay for this by QUANTUM NONLOCALTY.
I think that it is too high price for ignorance.

But, even by using the V\"axj\"o interpretation one should be
careful with the use of the projection postulate. In fact, {\bf
VXU2} also might be interpreted in two ways: von Neumann's like
and L\"uders-like. But, since this paper is solely based on the
Copenhagen interpretation, we do not want to go into details.

Other people (experts in theory of so called "quantum
instruments") pointed to me that they are well aware about
different forms of the projection postulate, see e.g. \cite{DV}--
\cite{BU}. And it is nothing new for them. However, they either
proceed in purely mathematical framework or even simply ignore the
principle physical difference between von Neumann's and
L\"uders'versions of the projection postulate. In the latter case
they even speak about von Neumann-L\"uders' postulate by
considering L\"uders' postulate as just a natural generalization
of von Neumann's one. Typically von Neumann's postulate is
considered as a "primitive" one which was "improved" by L\"uders.

\section{The role of the projection postulate in the EPR argument}

The role of the projection postulate in the EPR-considerations is
practically unknown (except of  a few experts in quantum
foundations). The main problem is that not so many people have
read the original EPR-paper \cite{EPR}. Even if one did this, it
was not careful reading - since it was easier to understand the
EPR-arguments from later books on QM. However the projection
postulate is the basis of the EPR-definition of an element of
reality.\footnote{From the very beginning we emphasize that the
EPR-arguments were against QM as a theoretical model (including
interpretational part). Thus the EPR story was not about "physical
elements of reality", but about their  theoretical counterparts in
the formalism of QM. We recall that axiomatization of QM was
performed by Dirac \cite{D} and von Neumann  \cite{VN}.
Measurement theory was completely formalized in \cite{VN}. EPR's
arguments are in fact about measurement theory. To be rigorous,
they should speak about theoretical counterparts of "elements of
reality" in von Neumann's axiomatic model. Unfortunately, EPR did
not do this precisely (as we shall see). Instead of speaking about
von Neumann's axiomatics, they criticized  a QM model which was
not rigorously formalized. I think that this absence of rigor was
the main root of the "EPR-paradox."}  Hence, its use (in fact,
misuse) is the main source of dilemma: either incompleteness or
nonlocality. We shall see that the right (von Neumann) application
of the projection postulate would not generate such a dilemma. In
particular, so called "quantum nonlocality" would not at all
appear  in discussion on completeness of QM (its Copenhagen
interpretation).

What was wrong in the EPR-considerations? The crucial point was
misuse of reduction of wave function in QM. By speaking about QM
one should pay attention both to its mathematical formalism and
its interpretation. The EPR consideration was not consistent
neither with the mathematical formulation (due to von Neumann \cite{VN})
 nor interpretation (due to Bohr \cite{BR}).

We now present the EPR-arguments in detail, since otherwise it
would be really impossible to criticize them: details are
extremely important. We remind the EPR viewpoint on elements of
reality:

\medskip

 ``If, {\it without in any way disturbing a system, we can predict with certainty
(i.e., with probability equal to unity) the value of a physical
quantity then there exists an element of physical reality
corresponding to this physical quantity.''}

\medskip

We emphasize that  the main part of the EPR paper \cite{EPR}
consists of considerations on description of reduction of the wave
function in QM. Their aim was to associate elements of reality
with elements of the theoretical model of QM. We recall that the
{\it EPR critique was against this model} (and not at all against
some real experimental designs). We shall see that EPR associated
their elements of reality with eigenfunctions of corresponding
self-adjoint operators. We now present their considerations on
reduction.

If $\psi$ is an eigenfunction of the operator $\widehat{A},$
\begin{equation}
\label{11.1}
 \psi^\prime\equiv\widehat{A}\psi= a \psi,
\end{equation}
where $a$ is a number, and so the physical quantity $A$ has with
certainty the  value $a$ whenever the particle is in the state
$\psi.$  By the criterion of reality, for a particle in the state
given by $\psi$ for which (\ref{11.1}) holds there is an element
of physical reality corresponding to the physical quantity $A.$
For example,
\begin{equation}
\label{11.2} \psi=e^{(i/\hbar) p_0 x},
\end{equation}
where $p_0$ is some constant number, and $x$ the independent
variable. Since the operator corresponding to the momentum of the
particle is
\begin{equation}
\label{11.3}
 \widehat p = \frac{\hbar}{i}\frac{\partial}{\partial
x},
\end{equation}
we obtain
\begin{equation}
\label{11.4} \psi^\prime =\widehat p \psi
=\frac{\hbar}{i}\frac{\partial}{\partial x}\psi  = p_0\psi.
\end{equation}
Thus in the state given by (\ref{11.2}) the momentum has certainly
the value $p_0.$ It thus has meaning to say that the momentum of
the particle in the state given by (\ref{11.2}) is real.

On the other hand, if (\ref{11.1}) does not hold we can no longer
speak of the physical quantity $A$ having a particular value. This
is the case, for example, with the coordinate of the particle. The
operator corresponding to it, say $\widehat q,$ is the operator of
multiplication by the independent variable. Thus
\begin{equation}
\label{11.5} \widehat q \psi=x\psi \not=a\psi.
\end{equation}
In accordance with quantum mechanics we can only  say that the
relative probability  that a measurement of the coordinate will
give a result lying between $a$ and $b$ is
\begin{equation}
\label{11.6} {\bf P}_\psi([a,b])=\int_a^b \psi\bar{\psi} d x
=\int_a ^b d x = b - a.
\end{equation}
Since this probability  depends upon the difference $b-a,$ we see
that all values of the coordinate are equally probable.

More generally, if the operators corresponding to two physical
quantities, say $A$ and $B$, do not commute, that is, if
$[\widehat{A}, \widehat{B}]= \widehat{A}\widehat{B} -
\widehat{B}\widehat{A} \not= 0,$ then the precise knowledge of one
of them precludes such a knowledge of the other. Furthermore, any
attempt to determine the latter experimentally will alter the
state of the system in such a way as to destroy the knowledge of
the first.

From this it follows that: either

a) {\it the quantum mechanical description of reality given by the
wave function is not complete;}

or

b) {\it when the operators corresponding to two physical
quantities do not commute the two quantities cannot have
simultaneous reality.}

For if both of them had simultaneous reality--and thus definite
values--these values would enter into the complete description,
according to the condition of completeness. If then the wave
function provided such a complete description of reality, it would
contain these values; these would be predictable.

By the Copenhagen interpretation of  quantum mechanics it is
assumed that the wave function does contain a complete description
of the physical reality of the system in the state to which it
corresponds.

Let us suppose that we have two systems $S_1$ and $S_2$ which we
permit to interact from the time $t=0$ to $t=T,$ after which time
we suppose that there is no longer any interaction between the two
parts. We further suppose that the states of the two systems
before $t=0$ were known. We can then calculate, with the help of
the Schr\"odinger equation, the state of the combined system $S_1
+S_2$ at any subsequent time; in particular, for any $t> T.$

Let us designate the corresponding wave function (calculated with
the aid of the Schr\"odinger equation) by $\Psi.$ This is the
function of the two variables $x_1$ and $x_2$ corresponding to the
systems $S_1$ and $S_2$ respectively, $\Psi=\Psi(x_1,x_2).$
 We
cannot, however, calculate the state in which either one of the
two systems is left after the interaction. This, according to
quantum mechanics, can be done with the help of the further
measurements by a process known as the {\it reduction of the wave
function.} Let us consider the essentials of this process.

Let $a_1,a_2,a_3,...$ be the eigenvalues of an operator
$\widehat{A}$ corresponding to some physical quantity $A$
pertaining to the system $S_1$ and $u_1(x_1), u_2(x_1),$
$u_3(x_1),...$ the corresponding eigenfunctions. Then $\Psi,$
considered as a function of $x_1,$ can be expressed as
\begin{equation}
\label{11.7} \Psi(x_1,x_2)= \sum_{n=1}^\infty  u_n(x_1)
\psi_n(x_2)
\end{equation}
Here the $\psi_n(x_2)$ are to be regarded merely as the
coefficients of the expansion of $\Psi(x_1,x_2)$ into a series of
orthogonal functions $u_n(x_1).$ Suppose now that the quantity $A$
is measured and is found to have the value $a_k.$ It is then
concluded that after the measurement the first system is left in
the state given by the wave function $u_k(x_1),$ and the second
system is left in the state given by the wave function
$\psi_k(x_2).$ This is the process of reduction of the wave
function; the wave function given by the infinite series
(\ref{11.7}) is reduced to a single term $ u_k(x_1) \psi_k(x_2).$

The set of functions $u_n(x_1)$ is determined by the choice of the
physical quantity $A.$ If, instead of this, we had chosen another
quantity, say $B,$ with the operator $\widehat{B}$ having the
eigenvalues $b_1, b_2, b_3,... $ and eigenfunctions
$v_1(x_1),v_2(x_1),v_3(x_1),... $ we should have obtained, instead
of (\ref{11.7}), the expansion
\begin{equation}
\label{11.8} \Psi(x_1,x_2)=\sum_{s=1}^\infty  v_s(x_1)
\phi_s(x_2),
\end{equation}
where $\phi_s$ are the new coefficients. If the quantity $B$ is
now measured and is found to have the value $b_r,$ we conclude
that after the measurement the system $S_2$ is left in the state
given by $\phi_r(x_2).$

Let us now go back to the consideration of the quantum state
$\Psi.$ As we have seen, as a consequence of two different
measurements performed upon the first system $S_1$ (for the
quantities $A$ and $B$) the second system may be left in states
with two different wave functions -- $\psi_k(x_2)$ and
$\phi_r(x_2)$. On the other hand,
 since at the time of measurement the two systems no longer
interact, no real change can take place in the second system as a
consequence of anything that may be done to the first system. This
is, of course, merely a statement of what is meant by the absence
of an interaction between the two systems. Thus {\it it is
possible to assign two different wave functions} (in our example
$\psi_k$ and $\phi_r$) {\it to the same reality} (the second
system after the interaction with the first).

Now, it may happen that the two wave functions $\psi_k$ and
$\phi_r$ are eigenfunctions of two non-commuting operators
corresponding to some physical quantities $P$ and $Q,$
respectively. That this may actually be the case can best be shown
by an example, see \cite{EPR}.

\section{On the logical scheme of the EPR argument}
\label{SSU}

1). EPR provided their own definition of "an element of
reality."We point out that it does not belong to the theoretical
model of QM. Hence they should map "elements of reality" onto some
conventional objects of the QM-model. EPR understood well that one
could not criticize one theoretical model by using notions from a
different model.

2). To perform such a task, EPR used the following consequence of
the projection postulate.  Let $A$ be a (self-adjoint) operator
representing  quantum observable. Let $\psi$ be its eigenvector.
So, (\ref{11.1}) holds. Then the value $A=a$ can be predicted with
certainty. It justifies association of EPR's elements of reality
with eigenvectors. Thus (at least some) elements of reality can be
represented by eigenvectors in the the QM-model. It is important
that any eigenvector represents an element of reality.

3). By {\it using the QM-model} EPR proved that one can assign to
the same  system  eigenfunctions corresponding to noncommuting
operators.

We shall criticize the last step of EPR's considerations.

\section{The von Neumann projection postulate}

In von Neumann's book \cite{VN} the cases of observables with {\it
nondegenerate and degenerate spectra} were sharply distinguished.
The post-measurement state is well defined (and given by the
corresponding eigenvector) only for observables with nondegenerate
spectra. Only in this case EPR might say that one could assign the
wave function with the physical system (after the measurement).
However, if spectrum is degenerate, then by the von Neumann
axiomatics of QM the post-measurement state is not determined.

Thus one could not assign the definite wave function with the
physical system (after measurement).

It is amazing that EPR did not pay attention to this crucial
point. I could not exclude that they even did not read von
Neumann's book. In their paper the projection postulate is applied
for observables with degenerate spectra, but in such a way as if
they were observables with nondegenerate spectra.

By considering
partial measurements on subsystems of composite systems one
immediately moves to the domain of degenerate measurements. Those
operators $A$ and $B$ considered by EPR have degenerate spectra.
Therefore by measuring e.g. $A$ one would not determine the state
of a composite system $S_1 + S_2.$ Hence, the state of $S_2$ is
not determined by $A$-measurement on $S_1.$ The wave function
$\psi_k (x_2)$ could not be assigned with $S_2.$ It is impossible to
proceed as EPR did at the very end of their general
considerations on measurements on composite systems. Since even
one wave function, $\psi_k (x_2),$ could not be assigned with
$S_2$, it is totally meaningless to write about assigning of {\it
two different wave functions to the same reality.}

\medskip

{\bf Conclusion.} {\it EPR did not prove that QM is incomplete.
They did mistake by assuming that by measurement of observable $A$
(respectively, $B$) on $S_1$ the linear combination (\ref{11.7})
(respectively, (\ref{11.8}))  is reduced to a single summand.}

\section{EPR is about precise correlations}

My correspondence with readers of preprint \cite{KHRV}
demonstrated that considerations of EPR on reduction of the wave
function (which were presented in section 2) have never been
discussed seriously. This part of EPR's paper (two of totally four
pages) is practically ignored. Instead of this, people have always
been concentrated on the last page of the paper containing the
discussion on precise correlations for the position and momentum.
As e.g. Elena Loubentz and Joachim Kupsch pointed out in E-mails
to me, the EPR paper is not about the projection postulate, but
about measurements for states with precise correlations. We remark
that mentioned  "presentation of the EPR without appealing to
reduction of wave function" can be found in the book of Ballentine
\cite{BL}, p.583-584. He really believes that he simplified the
EPR arguments and the he escaped using the notion of
reduction.\footnote{Hans de Raedt pointed out (in Email to me) to
Ballentine's presentation of the EPR views in \cite{BL}.}  We come
back to the original EPR argument.

The essence of the EPR conclusions is presented in short on page
780:

\medskip

"Returning now to the general case contemplated
in Eqs. (7) and (8), we assume that $\psi_k$ and $\phi_r$ are
indeed eigenfunctions of some non-commuting operators $P$ and $Q$,
corresponding to the eigenvalues $p_k$ and $q_r$, respectively.
Thus by measuring either $A$ or $B$ we are in a position to
predict with certainty, and without in any way disturbing the
second system, whether the value of the quantity $P$ (that is
$p_k)$ or the value of the quantity $Q$ (that is $q_r).$ In
accordance with our criterion of reality, in the first case we
must consider the quantity $P$ as being an element of reality, in
the second case the quantity $Q$ is an element of reality."

\medskip

As I understood, the last sentence has always been considered as
the very end of the story. However, (by some reason) EPR
continued:

\medskip

"But, as we have seen, both wave functions $\psi_k$ and $\phi_r$,
belong to the same reality."

\medskip

Opposite to the majority of readers of their paper or (and it was
more common) some texts about their paper, EPR were not able to
get the complete satisfaction via  producing  elements of reality
for the second particle via $A$ and $B$ measurements on the first
one. They had to come back to their rather long story (pages
788-789) on reduction of the wave function.

I think that this EPR's comeback to reduction is the crucial point
of their argument. Why did they need do this? I think that by the
following reason. It is impossible to associate simultaneously two
"experimental elements of reality" with $S_2$ on the basis of
measurement on $S_1,$ since (as everybody understood well) either
$A$ or $B$ measurement could be performed on $S_1$ (but not both
$A$ and $B$). Therefore EPR were able to associate with $S_2$ only
"theoretical elements of reality" represented by the wave
functions $\psi_k (x_2)$ and $\phi_r (x_2)$ - eigenfunctions of
the two non-commuting operators $P$ and $Q$ (for the second
particle).

And it was enough for their purpose, since they wanted to prove
incompleteness of QM as a {\it theoretical} model, see section
\ref{SSU}. Thus, although I have the great respect to the
contribution of Ballentine to quantum foundations, I do not think
that his viewpoint is correct. EPR were clever enough to restrict
their argument  to Ballentine's type considerations \cite{BL},
p.583-584. They did not do this just because they were not able to
approach their aim in this way.

\medskip

{\bf Conclusion.} {\it EPR were not able to proceed without
appealing to the projection postulate (with all consequences of
its misuse).}

\section{Refinement measurements}
However, according to von Neumann by obtaining a fixed value, say
$A=\alpha, $ for measurement on  $S_1,$ one does not determine
the state of $S_1 + S_2$ (and, hence, neither the state of $S_2$).

To determine the state of $S_1 + S_2,$ one should perform some
{\it refinement measurement.} In QM it is represented by an operator
commuting with $A \otimes I$ and eliminating
degeneration\footnote{Here $A: L_2 ({\bf R}^3) \to L_2 ({\bf
R}^3), A \otimes I: L_2 ({\bf R}^3) \otimes L_2 ({\bf R}^3) \to
L_2 ({\bf R}^3) \otimes L_2 ({\bf R}^3).$}. Since any operator of
the form $I \otimes C$ commutes with $A \otimes I,$ it is natural
to consider refinement observable corresponding to measurement on
$S_2.$ The position Q and momentum P operators considered by EPR
give examples of von Neumann's refinement measurements. Each of
them determine the state of $S_1 + S_2$ (and hence $S_2$)
uniquely.

Moreover, for any operator with degenerate spectrum its
measurement is ambiguous \cite{VN}. Thus in the EPR case
measurement of $A$ could not at all be considered as measurement on
$S_1+S_2.$ It is just measurement on $S_1.$

However, for EPR the story about so called EPR-states was not
simply the standard story about von Neumann's refinement
measurements.

\section{The EPR paper as the source of the idea about quantum nonlocality} At
the very end of their paper EPR discussed a problem which later
became known as the problem of {\it quantum nonlocality:}

\medskip

"One could
object to this conclusion on the grounds that our criterion of
reality is not sufficiently restrictive. Indeed, one would not
arrive at our conclusion if one insisted that two or more physical
quantities can be regarded as simultaneous elements of reality
{\it only when they can be simultaneously measured or predicted.}
On this point of view, since either one or the other, but not both
simultaneously, of the quantities $P$ and $Q$ can be predicted,
they are not simultaneously real. This makes the reality of $P$
and $Q$ depend upon the process of measurement carried out on the
first system, which does not disturb the second system in any way.
No reasonable definition of reality could be expected to permit
this."

\medskip

Later nonlocality was coupled to the von Neumann projection
postulate in the following way. To escape incompleteness of QM,
one should not assign the wave function $\psi_k (x_2)$ with $S_2$
before the $A$-measurement on $S_1.$ One might say that the
$A$-measurement on $S_1$ produces instantaneous action on $S_2$
and its state is collapsed into $\psi_k (x_2).$ For example, one
can find an example of such a reasoning in the paper of Alain
Aspect \cite{AS}.

This form of reasoning has  nothing to do with QM. By the same von
Neumann's projection postulate the state of $S_2$ is NOT
determined by measurement on $S_1.$ There is no even trace of
action at the distance!

\medskip

{\bf Conclusion.} {\it "Quantum nonlocality" appeared as a
consequence of misuse of the projection postulate. We also
emphasize that EPR considered quantum nonlocality as a totally
absurd alternative to their arguments in favor of incompleteness
of QM. }

\section{Nonlocality of the experiment design as opposed to EPR state
nonlocality}

\subsection{Quantum theory and joint measurements of compatible observables}
We have already discussed that from the QM-viewpoint (based on von
Neumann's axiomatics) the whole EPR story is about refinement
measurements for operators with degenerate spectra. It would be
useful to analyse (by using the conventional QM-framework) the
procedure of joint measurement of two compatible observables, say
$A$ and $Q: [A, Q]=0.$

The crucial point is that by von Neumann, to design joint
measurement of $A$ and $Q$, one should design measurement of {\it
third observable}, say $C$, such that $A= f(C)$ and $Q=g(C)$,
where $f, g: {\bf R} \to {\bf R}$ are some functions. In the EPR
case we want to have $C$ with nondegenerate spectrum and $A$ is
observable on $S_1$ and $Q$ on $S_2.$

Since $A$ and  $Q$ are measured in different domains of ${\it
spacetime},$ the design of measurement of $C$ should be nonlocal.
It is an extremely important point.

What does it mean "nonlocal design"?

In particular, it means that one should perform the {\it time
synchronization} between results of measurement of $A$ and $Q$. It
is important to be totally sure that clicks of the $A$-detector
(giving the result of measurement on $S_1$)  and the $Q$-detector
(giving the result of measurement on $S_2$) match each other. We
emphasize that in the real experimental setup for the EPR-Bohm
experiment for photon polarization, see e.g., \cite{S1},
\cite{S2}, such a time synchronization is really realized via the
nonlocal experimental design - via using the {\it time window.}
The time window constraint
$$
|t^A_i - t^Q_i|<\Delta
$$
is evidently nonlocal. We also point out to the {\it
synchronization of space frames.} Orientations of polarization
beam splitters are chosen in one fixed space frame (in the
complete accordance with Bohr's ideology \cite{BR}).

\subsection{The EPR state nonlocality}

If one proceeds with so called quantum nonlocality induced by the
misuse of the projection postulate, then he should take such a
nonlocality very seriously. It would be real physical nonlocality
of states. We again recall that EPR considered such a nonlocality
as totally absurd.

\medskip

{\bf Conclusion.} {\it The correct application of the projection
postulate implies the nonlocal experimental design of the EPR-type
experiments; in particular, the time synchronization (e.g., via
the time window) as well as the choice of the fixed space frame.
This experimental design nonlocality has nothing to do with so
called "quantum nonlocality".}

\section{Bohr's reply to Einstein}
It is typically emphasized that Bohr's reply \cite{BR} is very
difficult for understanding. I totally agree with such a common
viewpoint. I was able to understand Bohr only on the basis of
previous considerations on the role of the projection postulate in
the EPR considerations. Unfortunately, in Bohr's reply there was
no even trace of von Neumann's axiomatization of QM \footnote{I
strongly suspect that neither Einstein nor Bohr had read von
Neumann's book at that time.}. Consequently Bohr did not pay any
attention to the role of the projection postulate in the EPR
considerations. He missed the EPR-trick with assigning to $S_2$
two wave functions, $\psi_k (x_2)$ and $\phi_r (x_2),$ which are
eigenfunctions of two noncommutative observables, say $P$ and $Q.$
It is very important in the EPR considerations that these wave
functions and not measurements by themselves represent "elements
of reality" in QM (as a theoretical model). Thus, instead of
analyzing this tricky point in the EPR paper, Bohr proceeded in
the purely experimental framework. He simply recalled his ideas on
complementarity of various measurement setups in relation to the
EPR-considerations. In short his message was that since one could
not combine two measurement setups for $S_1$ related to
incompatible quantities, it is impossible to assign two
corresponding elements of reality to $S_2.$ Bohr concluded that
the EPR notion of an element of reality was ambiguous.

The problem was that EPR "proved" that QM is incomplete as {\it a
theoretical model}, but Bohr replied by supporting his old thesis
that QM is complete as {\it an experimental methodology}. It seems
that the resulting common opinion was not in favor of Bohr's
reply. And it is clear why. If EPR really were able to prove that
the formalism of QM implies assigning to $S_2$ of two wave
functions, $\psi_k (x_2)$ and $\phi_r (x_2),$ corresponding to two
noncommuting operators $Q$ and $P$, I would (and I was!) on their
side. The point (presented in this paper)  is that they were not
able to do this by using the QM formalism in the proper way.

\medskip

{\bf Conclusion.} {\it Bohr's reply in spite correctness of his
arguments, did not contain the analysis of the real roots of the
"EPR paradox". It induced a rather common impression that EPR's
argument is not trivially reduced to the old problem of
complementarity. It was commonly accepted that the only
possibility to escape assigning "elements of reality"
corresponding to incompatible observables to the same particle is
to accept quantum nonlocality.}

\section{Concluding remarks}

It seems that the "EPR-paradox" was  finally resolved in this
paper. I hope that it would stimulate people to look for various
ways beyond QM.  By von Neumann's axiomatics of QM \cite{VN} the
notion of measurement of observable $A$ with degenerate spectrum
is ambiguous. It is well defined only via refinement measurement
given by observable $C$ with nongenerate spectrum such that
$A=f(C).$ Since any observable $A$ on the subsystem $S_1$ of a
composite system $S=S_1 + S_2$ has degenerate spectrum in the
tensor Hilbert space of $S$-states, it is totally meaningless to
discuss (as EPR did) its measurement without fixing a refinement
measurement on $S_2.$ If such a refinement is not fixed from the
very beginning, then $A$-measurement has nothing to do with
measurements on the composite systems $S.$ It could not change the
$S$-state and, hence, the $S_2$-state. Bohr's reply \cite{BR} to
Einstein could be interpreted in the same way. Thus the EPR-attack
against QM was not justified. Unfortunately, this attack was the
source of naive Einsteinian realism (assigning to the same system
$S_2$ of two wave functions $\psi_k (x_2)$ and $\phi_r(x_2)$
corresponding to noncommutative operators) and quantum
nonlocality. We also point out to practically unknown fact that so
called EPR states were studied in detail by von Neumann \cite{VN},
pp. 434-435. But he was able to proceed without assigning two wave
functions (corresponding to noncommuting operators) to the same
system. Consequently, no traces of  incompleteness of QM or its
nonlocality could be found in \cite{VN}.

Finally, we remark that recently Bell-type inequalities for tests
of compatibility of {\it nonlocal realistic models} with quantum
mechanics were derived, see Legget \cite{Legget}. They were
generalized and tested experimentally by Gr\"oblacher et al.
\cite{Grob}. The conclusion of these theoretical and experimental
studies is that the condition of nonlocality which was considered
by Bell (of course, under the influence of EPR) plays a subsidiary
role. It was proven that naive EPR-realism is incompatible with
experimental data (and this fact has no relation to the EPR-Bell
idea of nonlocality). It is an experimental confirmation that the
analysis of the EPR-arguments performed in the present paper is
correct. These arguments were wrong from the very beginning.

I would like to thank A. Majewski, K. Hess, A. Plotnitsky, E.
Loubentz, J. Kupsch,  H. de Raedt, V. Manko and O. Manko for
critical comments on my preprint \cite{KHRV} and A. Grib, R. Gill,
M. Kupsczynski, A. Holevo, Yu. Bogdanov,  Yu. Ozhigov for critical
comments on this preprint.

\end{document}